\documentclass[aps,prc,10pt,showpacs,showkeys,twocolumn,superscriptaddress,groupedaddress]{revtex4-1}

\usepackage[english]{babel}
\usepackage{indentfirst}
\usepackage{amssymb}
\usepackage{braket}
\usepackage{bm}
\usepackage{amsxtra}
\usepackage{amsmath}
\usepackage{supertabular}
\usepackage{multirow}
\usepackage{color}
\usepackage [mathcal]{eucal}
\usepackage{graphicx}
\usepackage{graphics}
\usepackage{lipsum}

\def\n4lo{$\mathrm{N}^4\mathrm{LO}$}
\def\n3lo{$\mathrm{N}^3\mathrm{LO}$}

     \def\chiral4lo{$\mathrm{N}^4\mathrm{LO}$}

\begin{document}

\noindent
\title{Elastic Antiproton-Nucleus Scattering from Chiral Forces}

\author{Matteo Vorabbi$^{1,2}$}
\author{Michael Gennari$^{2,3}$}
\author{Paolo Finelli$^{4}$}
\author{Carlotta Giusti$^{5}$}
\author{Petr Navr\'atil$^{2}$}

\affiliation{$~^{1}$National Nuclear Data Center, Bldg. 817, Brookhaven National Laboratory, Upton, New York 11973-5000, USA
}

\affiliation{$~^{2}$TRIUMF, 4004 Wesbrook Mall, Vancouver, British Columbia, V6T 2A3, Canada
}

\affiliation{$~^{3}$University of Victoria, 3800 Finnerty Road, Victoria, British Columbia V8P 5C2, Canada
}

\affiliation{$~^{4}$Dipartimento di Fisica e Astronomia, 
Universit\`{a} degli Studi di Bologna and \\
INFN, Sezione di Bologna, Via Irnerio 46, I-40126 Bologna, Italy
}

\affiliation{$~^{5}$Dipartimento di Fisica,  
Universit\`a degli Studi di Pavia and \\
INFN, Sezione di Pavia,  Via A. Bassi 6, I-27100 Pavia, Italy
}

\date{\today}


\begin{abstract} 

Elastic scattering of antiprotons off $^{4}$He, $^{12}$C, and $^{16,18}$O is described for the first time with a consistent
microscopic approach based on the calculation of an optical potential (OP) describing the antiproton-target interaction. The OP is derived using
the recent antiproton-nucleon ($\bar{p}N$) chiral interaction to calculate the $\bar{p}N$ $t$ matrix, while the target densities are computed with the {\it ab initio} no-core shell
model using chiral interactions as well. Our results are in good agreement with the existing experimental data and the results computed at different chiral orders of
the $\bar{p}N$ interaction display a well-defined convergence pattern.
\end{abstract}


\maketitle


With the Facility for Antiproton and Ion Research (FAIR) construction under way~\cite{fair} and the recent antiProton Unstable Matter Annihilation (PUMA) 
proposal \cite{Obertelli:2622466, Aumann:2691045}, scientific interest in new experiments on antiproton scattering off nuclear targets
(nucleons and nuclei) will experience a renaissance. 

In the past, there has been a lot of activity in the antiproton physics at the Low Energy Antiproton Ring (LEAR) at CERN as
well as at KEK in Japan and Brookhaven National Laboratory (BNL) in the USA. At LEAR, in particular, several measurements of cross sections have been made for antiproton
elastic and charge-exchange scattering reactions at antiproton momenta in the
range~$100$ MeV/c $\le p \le 2$ GeV/c \cite{annurev.ns.41.120191.001251,nla.cat-vn777045,nla.cat-vn257837,annurev.ns.38.120188.000435}.

The dominant feature of antiproton-proton scattering at low energies, i.e., the energy region on which our Letter is focused, is the annihilation process that, due to its large
cross section, greatly reduces the probability of rescattering processes. Antiproton-nucleus ($\bar{p}A$) scattering is thus likely to be described by simple reaction mechanisms
without the complication of multiple scattering processes, which makes it a very {\it clean} method to study nuclear properties. 
In fact, the pronounced diffraction structure of the differential cross sections (in contrast with elastic proton scattering) is commonly interpreted as a consequence of the role played by
the strong absorptive potential driven by the annihilation of nucleons and antinucleons. 
Antiproton absorption is surface-dominated \cite{annurev.ns.38.120188.000435,DOVER199287,ADACHI1987461} and is strongly sensitive to nuclear radii. 
The exchange mechanism and the antisymmetrization between the projectile and the target constituents are not relevant in the $\bar{p}A$ interaction, while
the role played by the three-body forces involving an antiproton and two nucleons ($\bar{p}NN$) still remains an open question.


From the theoretical point of view, the description of antiproton-nucleon ($\bar{p}N$) processes was mainly based on long-range meson exchanges, with the addition of
phenomenological models for annihilation contributions. Several approaches have been proposed over the last forty years. One of the most successful potentials is the model
proposed by Dover and Richard \cite{PhysRevC.21.1466,PhysRevC.25.1952} who were inspired by the Paris potential. Other antinucleon-nucleon ($\bar{N}N$) interactions, based
on the meson theory, were also derived~\cite{PhysRevC.39.761,PhysRevC.79.054001}, where the $\bar{N}N$ potential of Ref.~\cite{PhysRevC.79.054001} was used to
study $\bar{p}A$ quasibound states~\cite{HRTANKOVA201845}. A more general approach~\cite{PhysRevC.86.044003} was also employed to provide a partial-wave analysis
of antiproton-proton data.
A similar situation is found for $\bar{p}A$ scattering processes. In the 80s, several nonrelativistic and relativistic calculations were performed with different
approaches which made use of an optical potential (OP) \cite{hodgson1963} but required some phenomenological input. A summary of all these calculations can be found
in Ref.~\cite{Heiselberg1989}. Even in recent years new phenomenological models have been
presented~\cite{PhysRevC.54.332,GAITANOS2015181,FRIEDMAN2015101,LARIONOV2017450}.

Because of the tremendous advances in computational techniques achieved in the past decades, it is now possible to compute the OP for $\bar{p}A$ scattering in a fully
microscopic and consistent way. The purpose of this Letter is to construct the first fully microscopic OP for elastic $\bar{p}A$ scattering using the most recent techniques in nuclear
physics, in particular, the application of chiral $\bar{p}N$ potentials combined with nuclear densities obtained from {\it ab initio} calculations with chiral two- ($NN$) and
three-nucleon ($3N$) interactions. The results for the elastic differential cross sections produced with our OP will be then tested against the existing experimental data.
For such a purpose, we adopt a scheme analogous to that employed in Ref.~\cite{Gennari:2017yez}, where a microscopic OP for proton-nucleus ($pA$) elastic scattering has been
derived within the Watson multiple scattering theory~\cite{Riesenfeld:1956zza} at the first order term of the spectator expansion~\cite{PhysRevC.52.1992} and assuming the impulse
approximation.
Recently, interest in the microscopic calculation of the OP for nucleon-nucleus ($NA$) processes produced several new papers and a very recent review can be found in
Ref.~\cite{Dickhoff2018}. Here we mention the work of Burrows {\it et al.}~\cite{Burrows:2018ggt}, which improved the calculation of the OP including the coupling between the target
nucleon and the residual nucleus, the work of Arellano and Blanchon~\cite{Arellano2018} on the irreducible nonlocality of the OP, the work of
Whitehead {\it et al.}~\cite{Whitehead:2018bfs} based on the calculation of the nucleon self-energy within the framework of the improved local density approximation, and
the work of Kohno~\cite{PhysRevC.98.054617} on the Pauli rearrangement potential.

In the present work the OP is computed in momentum space as
\begin{equation}\label{fullfoldingop}
\begin{split}
U ({\bm q},{\bm K}; E ) &= \sum_{N=p,n} \int d {\bm P} \; \eta ({\bm q},{\bm K},{\bm P}) \\
&\times t_{\bar{p}N} \left[ {\bm q} , \frac{1}{2} \left( \frac{A+1}{A} {\bm K} + \sqrt{\frac{A-1}{A}} {\bm P} \right) ; E \right] \\
&\times \rho_N \left( {\bm P} + \sqrt{\frac{A-1}{A}} \frac{{\bm q}}{2} , {\bm P} - \sqrt{\frac{A-1}{A}} \frac{{\bm q}}{2} \right) \, ,
\end{split}
\end{equation}
where ${\bm q}$ and ${\bm K}$ represent the momentum transfer and the average momentum, respectively. Here ${\bm P}$ is an integration variable, $t_{\bar{p}N}$ is the $\bar{p}N$
free $t$ matrix and $\rho_N$ is the one-body nuclear density matrix. The parameter $\eta$ is the M\o ller factor, which imposes the Lorentz invariance of the flux when we pass from
the $\bar{p}A$ to the $\bar{p}N$ frame in which the $t$ matrices are evaluated. Finally, $E$ is the energy at which the $t$ matrices are evaluated and it is fixed at one-half 
the kinetic energy of the incident antiproton in the laboratory frame. 

The calculation of Eq.(\ref{fullfoldingop}) requires two basic ingredients: the $\bar{p}N$ scattering matrix and the one-body nuclear density of the target. The calculation of the
density matrix is performed using the same approach followed in Ref.~\cite{Gennari:2017yez}, where one-body translationally invariant (trinv) densities were computed within the
{\it ab initio} no-core shell model~\cite{BARRETT2013131} (NCSM) approach using $NN$ and $3N$ chiral interactions as the only input.
The NCSM method is based on the expansion of the nuclear wave functions in a harmonic oscillator basis and it is thus characterized by the harmonic oscillator
frequency $\hbar \Omega$ and the parameter $N_{max}$, which specifies the number of  nucleon excitations above the lowest energy configuration allowed by the Pauli principle.
In the present work we used the $NN$ chiral interaction developed by Machleidt {\it et al.}~\cite{Entem:2014msa,Entem:2017gor} up to the fifth order (N$^4$LO) with
a cutoff $\Lambda = 500$ MeV.
In addition to the $NN$ interaction, we also employed the $3N$ force to compute the one-body densities of the target nuclei. We adopted the $3N$ chiral
interaction derived up to third order (N\textsuperscript{2}LO), which employs a simultaneous local and nonlocal regularization with the cutoff values of
$650$ and $500$ MeV, respectively~\cite{Navratil2007,Gysbers2019}. The interaction is also renormalized using the similarity renormalization group (SRG) technique that
evolves the bare interaction at the desired resolution scale $\lambda_{\mathrm{SRG}}$ and ensures a faster convergence of our calculations. The densities have been computed
using $\hbar \Omega = 20$ MeV and $N_{max} = 14$ for $^4$He and $\hbar \Omega = 16$ MeV and $N_{max} = 8$ for $^{12}$C and $^{16,18}$O.
For all these calculations we always adopted $\lambda_{\mathrm{SRG}} = 2.0$ $\mathrm{fm}^{-1}$. Finally, the importance-truncated NCSM
basis~\cite{PhysRevLett.99.092501,PhysRevC.79.064324} was used for the $^{12}$C and $^{16,18}$O calculations at $N_{max} = 8$. We refer the reader to
Ref.~\cite{Gennari:2017yez} for all the details about the calculation of the densities and the removal of the center-of-mass contributions.

The same $NN$ interaction was used in Ref.~\cite{Gennari:2017yez} to compute the {\it pA} scattering matrix. 
The $\bar{p}N$ interaction is different from the
proton-nucleon ($pN$) one and in the present case it is not possible to compute the $t_{\bar{p}N}$ matrix with the same potential adopted for the calculation of the density.
For such calculation we use the first $\bar{p}N$ interaction at the next-to-next-to-next-to-leading order (\n3lo) in chiral perturbation theory (ChPT)
recently derived by Dai, Haidenbauer, and Mei\ss{}ner \cite{Dai2017}. In recent years, approaches based on ChPT had a great success, especially in the $NN$
sector \cite{Entem:2014msa, Entem:2017gor, Epelbaum:2014sza, Epelbaum:2014efa}. They are able to include symmetries and symmetry-breaking patterns of low energy QCD
and, at the same time, provide a reliable framework to express the $NN$ force in terms of a series of pion-exchange and contact interaction terms.
Two-body and many-body contributions naturally arise from the same prescriptions. The $NN$ reaction matrix is obtained solving a regularized Lippmann-Schwinger equation
for the bare $NN$ potential. We refer the reader to Refs.~\cite{bernard_ChPT_1995,VanKolck1999337} for a complete survey of ChPT and to
Refs.~\cite{Epelbaum:2008ga,Machleidt:2011zz} for the recent developments. Higher-order corrections to Eq. (\ref{fullfoldingop}) are very difficult to
estimate \cite{PhysRevC.51.1418}, in particular in a consistent picture along the chiral expansion of the $NN$ potential, but surely deserve future studies.

\begin{figure*}[t]
\begin{center}
\includegraphics[scale=0.65]{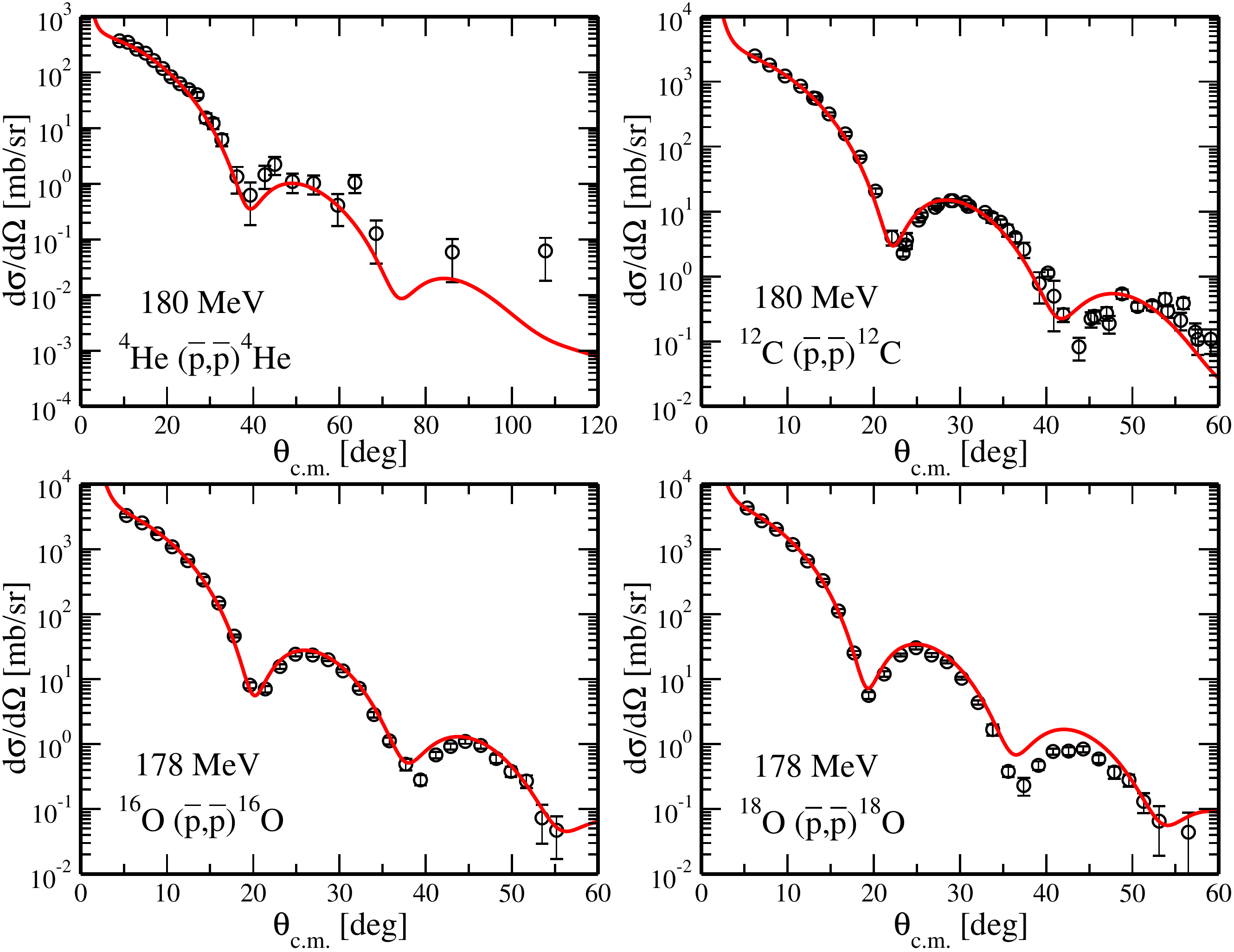}
\caption{Differential cross sections as a function of the center-of-mass scattering angle for elastic antiproton scattering off different target nuclei.
The results were obtained using Eq.(\ref{fullfoldingop}), where the $t_{\bar{p}N}$ matrix is computed with the $\bar{p}N$ chiral interaction of Ref.~\cite{Dai2017} and the
one-body trinv nonlocal density matrices are computed with the NCSM method using
two-~\cite{Entem:2017gor} and three-nucleon~\cite{Navratil2007,Gysbers2019} chiral interactions.
Experimental data from Refs.~\cite{Batusov:1990ge,GARRETA198464,BRUGE198614}.
\label{fig_cross_sections} }
\end{center}
\end{figure*}

In comparison with conventional $NN$ scattering, some issues must be addressed in the case of $\bar{N}N$ scattering. The main difference is that in the $\bar{N}N$ case the
annihilation channel is available because the total baryon number is zero. For low-momentum protons, elastic $\bar{p}N$ scattering requires a higher number of partial waves
compared to the $pN$ counterpart. All phase shifts are complex because of the annihilation process and both isospin 0 and 1 contribute in each partial
wave~\cite{PhysRevC.43.1610}. As a consequence, a treatment of $\bar{p}N$ scattering is intrinsically more complicated than the usual $NN$ system.

A conventional way to relate the $NN$ interaction to the $\bar{N}N$ counterpart is $G$ parity, i.e., a combination of charge conjugation and rotation in isospin space~\cite{Dai2017}.
It connects the pion-exchange physics, so even in the $\bar{N}N$ case the long-range physics is completely determined by chiral dynamics. In Ref.~\cite{Dai2017}, Dai {\it et al.}
developed a $\bar{p}N$ potential at \n3lo in analogy with the corresponding $NN$ potential presented in Refs.~\cite{chiralepelbaum_n3lo,Epelbaum:2014sza,Epelbaum:2014efa}, 
with the same power counting and a regularization scheme in the coordinate space. It seems that such a local scheme could avoid problems with the long-range part of the
interaction due to pion exchange that, of course, should not be affected by any regularization procedure.
We are aware of the many 
theoretical aspects beyond the regularization procedures (see Ref. \cite{hammer2019nuclear} and references therein) 
and more studies will be needed in the future.
In Ref. \cite{Dai2017}, five different potentials are provided with different values of the 
coordinate space cutoff $R$, that reproduce with almost the same quality the $\bar{N}N$ phase shifts. 
In the present work we employ the $R=0.9$ fm version.


\begin{figure}[t]
\begin{center}
\includegraphics[scale=0.35]{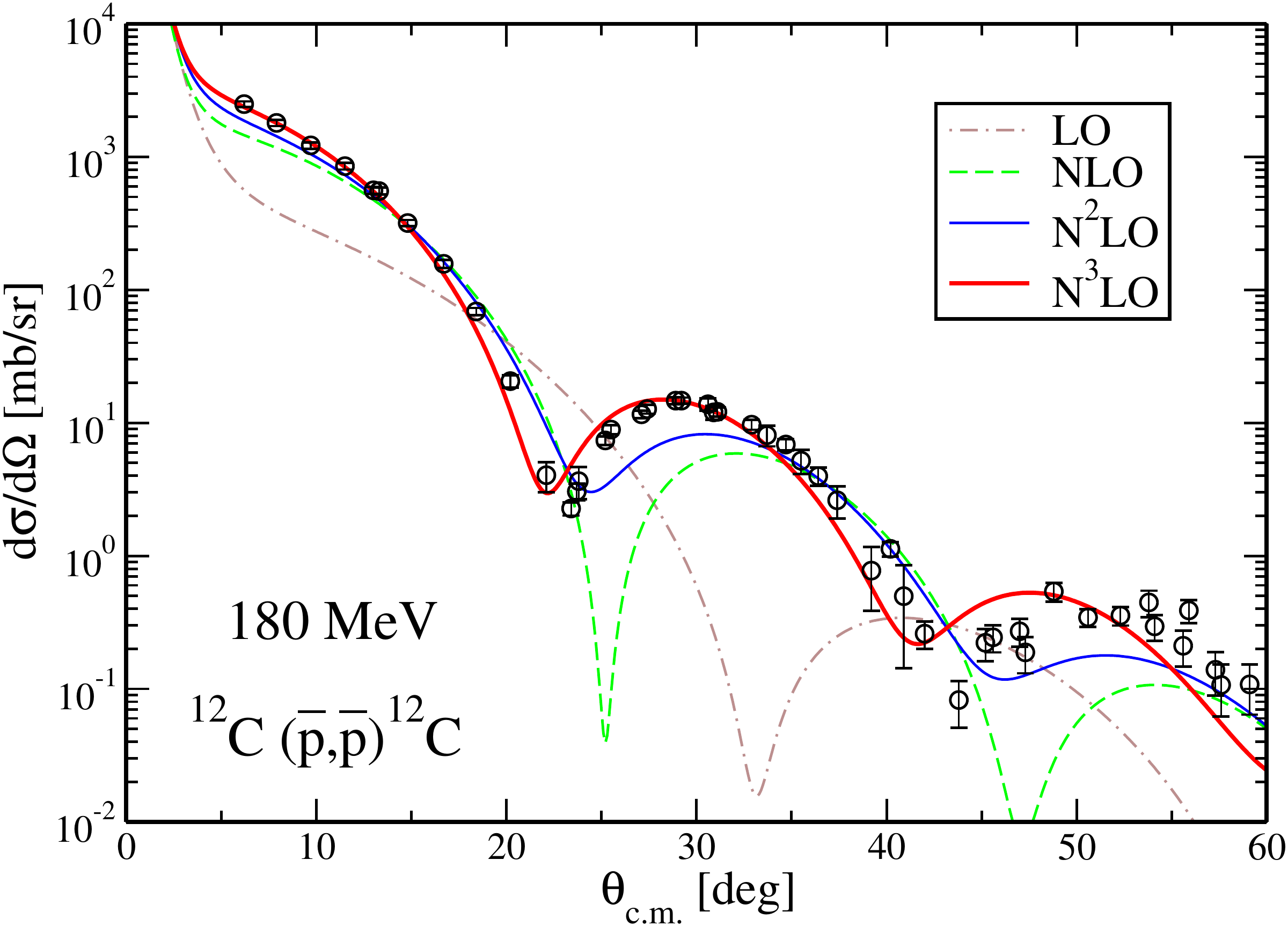}
\caption{Differential cross section as a function of the center-of-mass scattering angle for elastic antiproton scattering off $^{12}$C at 180 MeV, computed at different
chiral orders. Experimental data from Ref.~\cite{GARRETA198464}.
\label{fig_C12_order} }
\end{center}
\end{figure}

\begin{figure}[t]
\begin{center}
\includegraphics[scale=0.35]{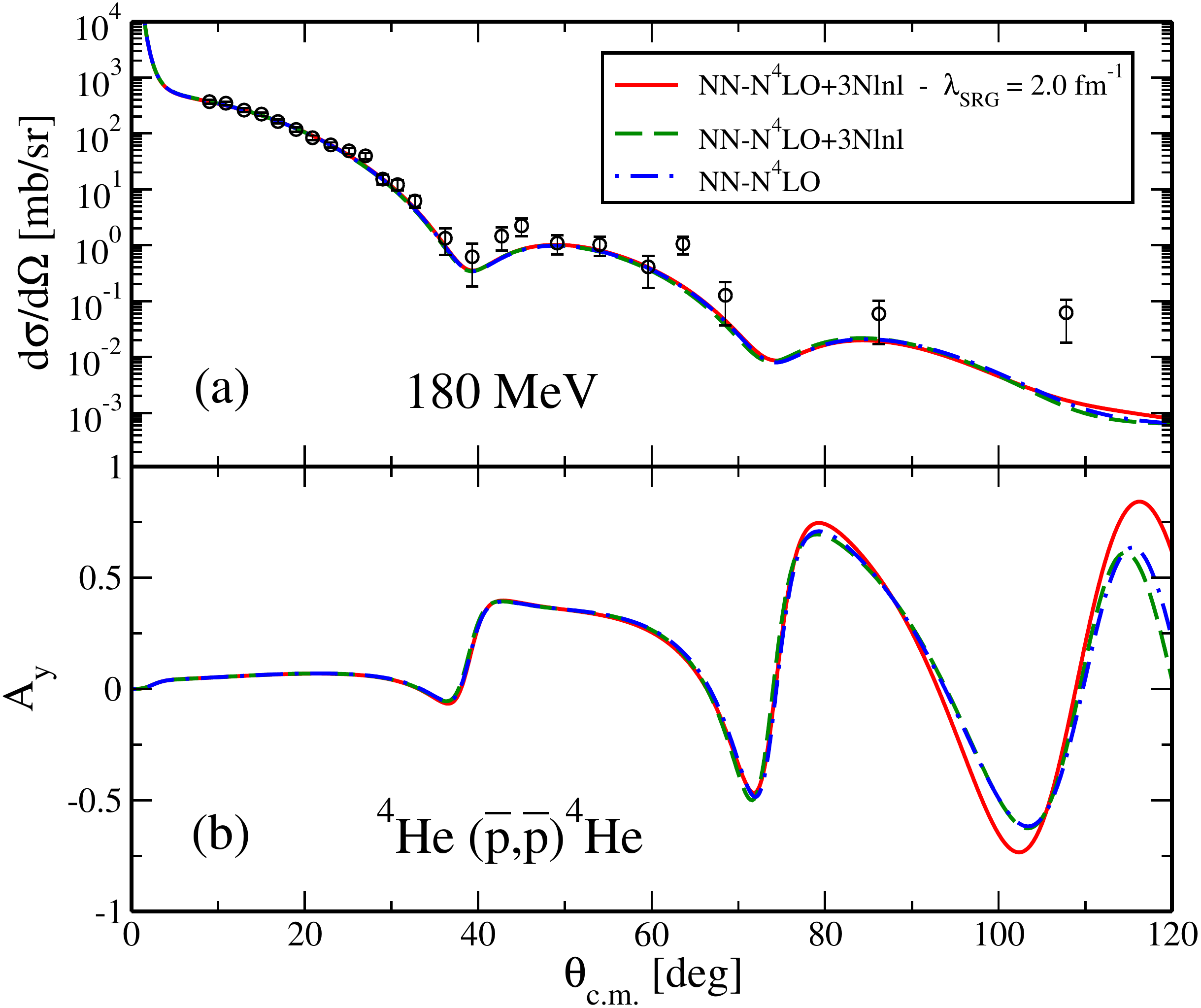}
\caption{Differential cross section (a) and analyzing power (b) as functions of the center-of-mass scattering angle for elastic antiproton scattering off $^{4}$He at 180 MeV.
The solid line represents the same result displayed in Fig.~\ref{fig_cross_sections}, the dashed line has been obtained with the target density computed without the SRG procedure,
while the dash-dotted line has been obtained with the target density computed with only the $NN$ interaction and without the SRG procedure. We always used the same values of
$N_{max}$ and $\hbar \Omega$.  Experimental data from Ref.~\cite{GARRETA198464}.
\label{fig_4He_comp} }
\end{center}
\end{figure}

\begin{figure}[t]
\begin{center}
\includegraphics[scale=0.35]{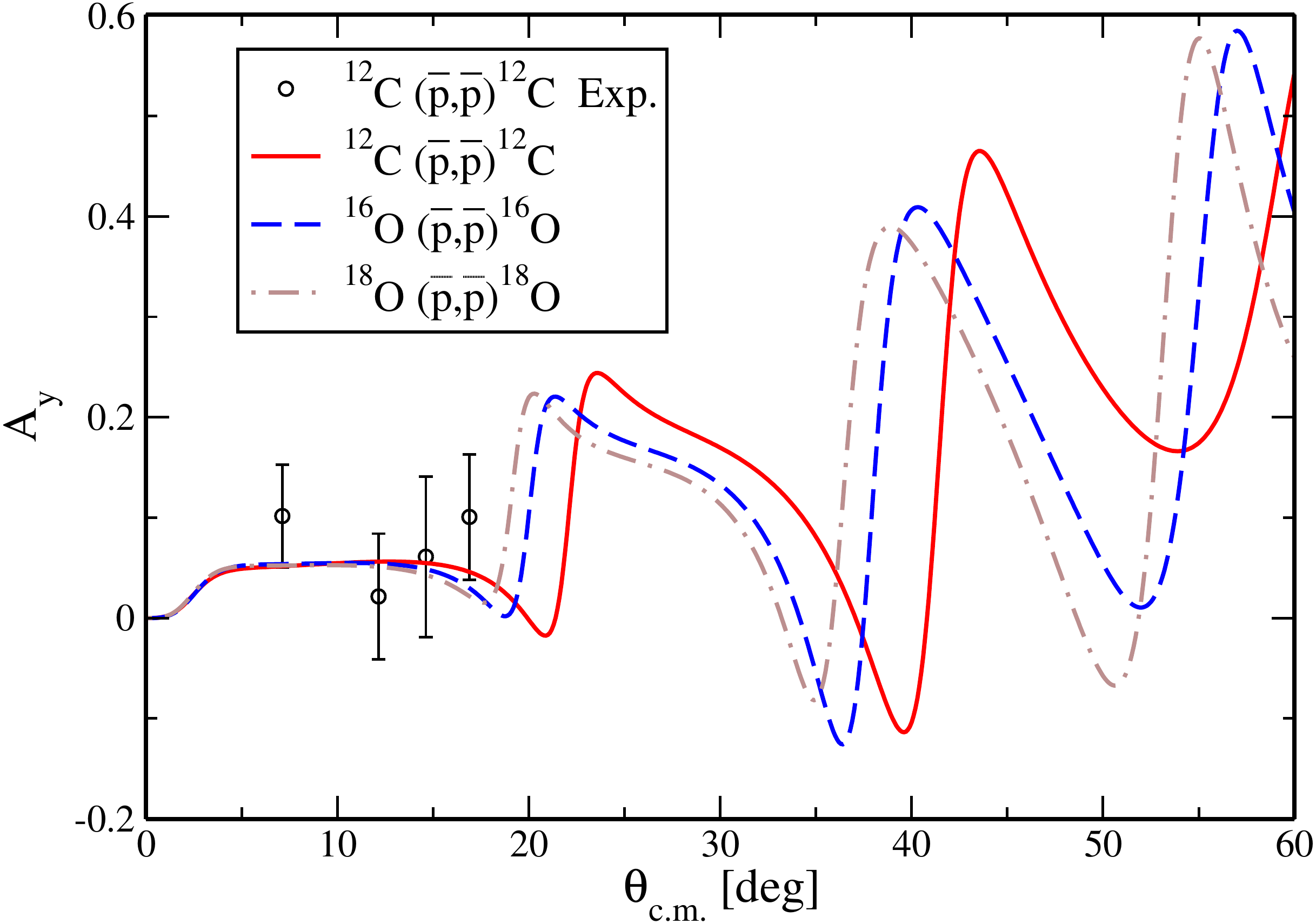}
\caption{Analyzing power as function of the center-of-mass scattering angle for elastic antiproton scattering off $^{12}$C and $^{16,18}$O computed at the same
energies and with the same inputs of Fig.~\ref{fig_cross_sections}.
\label{fig_Ay} }
\end{center}
\end{figure}

In Fig.~\ref{fig_cross_sections} our results for the differential cross sections of elastic antiproton scattering off $^4$He and $^{12}$C, computed at the antiproton laboratory
energy of 180 MeV, and $^{16,18}$O  at 178 MeV are presented and compared with the experimental data.
Our model provides a very good description of the data for all the target nuclei considered. In particular, it is remarkable the agreement in correspondence of the first minimum of
the diffraction pattern for all the targets and the general reproduction of the data for $^{18}$O, since this is an {\it sd} nucleus and is on the borderline of applicability of the NCSM.

One of the advantages of using a $NN$ or a $\bar{N}N$ interaction in the ChPT scheme is the ability to estimate the theoretical error associated with the truncation of the potential
at a certain order of the chiral expansion. In Fig.~\ref{fig_C12_order} we display the convergence pattern of the differential cross section for the $^{12}$C$(\bar{p},\bar{p}) ^{12}$C
reaction computed at different chiral orders. For a consistent comparison, all the calculations have been performed with the $\bar{p}N$ and $NN$ interactions at the same order
in the chiral expansion. For the calculation of the density at N$^2$LO and N$^3$LO we included the $3N$ force at N$^2$LO with the couplings $c_\mathrm{D}$ and $c_\mathrm{E}$
constrained to the triton half-life and binding energy. This produced two more fits of these parameters~\cite{quaglioni}, different from those employed with the $NN$ N$^4$LO
interaction, to be used with the $NN$ interaction at the same chiral order. All these results are displayed in Fig.~\ref{fig_C12_order}. As can be seen in the figure, at the leading order
(LO) the calculated cross section is in clear disagreement with data and has a minimum at about 33$^\circ$, which is more than 2 orders of magnitude lower than the experimental
one, which is positioned at about 23$^\circ$. A bit better result is obtained at NLO, where the first minimum is shifted towards smaller angles but the agreement with the experimental
cross section is still poor. At N$^2$LO the minimum is increased by about 2 orders of magnitude, close to the experimental value, but in comparison with the experimental cross
section the calculated cross section is shifted towards larger angles and the agreement with data remains poor. Only at the N$^3$LO the first minimum is well reproduced and the
general agreement with data is quite good. It is interesting to note how the differences between the results at different orders decrease going from LO to N$^3$LO, which reflects the
improvement and confirms a well-defined convergence pattern.
Similar results were found in Refs.~\cite{Vorabbi:2015nra,Vorabbi:2017rvk}, where a similar analysis was performed for $pA$ elastic
scattering using several chiral $NN$ interactions at $\mathrm{N}^3\mathrm{LO}$ and \chiral4lo.
The conclusion is that, for energies around 200 MeV, a good description of the experimental data is obtained with $NN$ or $\bar{N}N$ interactions up to at
least $\mathrm{N}^3\mathrm{LO}$. However, the choice of a different fitting procedure~\cite{PhysRevLett.110.192502} can produce an interaction capable to describe the
experimental data already at $\mathrm{N}^2\mathrm{LO}$, as recently showed in Ref.~\cite{Burrows:2018ggt} for the $NA$ case.

All the results presented so far were obtained with target densities computed using $NN$ and $3N$ interactions renormalized via the SRG. To assess the impact of the SRG
procedure in our calculations, we display in Fig.~\ref{fig_4He_comp} the results for the differential cross section and analyzing power for $^{4}$He computed with the bare $NN$ and
$3N$ interactions and the same values of $N_{max}$ and $\hbar \Omega$. The results are also compared with the ones in Fig.~\ref{fig_cross_sections}. As can be inferred from the
figure, the resulting densities produce the same results with minor differences at large scattering angles. Unfortunately, this is the only fully consistent calculation that we can perform
at the moment, since, in general, the usage of the bare interaction requires higher values of the $N_{max}$ parameter for a complete convergence of the structure calculations and
this is computationally prohibitive for heavier systems like carbon or oxygen.

Finally, in Fig.~\ref{fig_Ay} we display our predictions for the analyzing power of $^{12}$C and $^{16,18}$O, computed at the same energies and with the same inputs of
Fig.~\ref{fig_cross_sections}. We also show the only available experimental data \cite{1985PhLB..155..437B} obtained on carbon targets as part of the LEAR run of experiments.
The measured asymmetries are small, statistically compatible with zero, and suggest smaller polarisation parameters than those predicted by some $\bar{N}N$ phenomenological
potential models (see Fig. 11 of Ref. \cite{Heiselberg1989}). Our predictions, on the other hand, are consistent with measurements.


In summary, a fully microscopic OP for $\bar{p}A$ scattering has been derived within the Watson multiple scattering theory using the $\bar{N}N$, $NN$, and
$3N$ chiral interactions as the only input for our calculations. The new $\bar{N}N$ interaction derived up to $\mathrm{N}^3\mathrm{LO}$ has been used in our
calculations to obtain the $t_{\bar{p}N}$ scattering matrix needed in Eq.(\ref{fullfoldingop}). We tested our OP in comparison with the available experimental data for antiproton
elastic scattering off $^{4}$He, $^{12}$C, and $^{16,18}$O. Our results are in good agreement with data and are able to reproduce the correct angular position of the diffraction
minima. The OP has been also computed using the $\bar{p}N$ interaction at lower orders in
the chiral expansion to test the convergence of our results. As obtained in previous $pA$ calculations, also in this case for a good description of the data it is mandatory to use an
interaction derived at least up to $\mathrm{N}^3\mathrm{LO}$. As a concluding remark, we mention that at this stage new questions arise about the importance of $\bar{p}NN$
interactions in both structure and reaction calculations.


The authors are deeply grateful to L.-Y. Dai, J. Haidenbauer and U. Mei\ss{}ner who kindly provided the potentials derived in Ref. \cite{Dai2017}.
We are also grateful to A. Rotondi, A. Larionov, and H. Lenske for providing the antiproton-nucleus experimental data.
Finally, we would like to thank S. Quaglioni for providing the $c_D$ and $c_E$ parameters used in the convergence analysis.
The work at Brookhaven National Laboratory was sponsored by the Office of Nuclear Physics, Office of Science of the U.S. Department of Energy under
Contract No. DE-AC02-98CH10886 with Brookhaven Science Associates, LLC.
The work at TRIUMF was supported by the NSERC Grant No. SAPIN-2016-00033. TRIUMF receives federal funding via a contribution agreement with the National
Research Council of Canada.
Computing support came from an INCITE Award on the Titan supercomputer of the Oak Ridge Leadership Computing Facility (OLCF) at ORNL, from Westgrid and
Compute Canada.

%

\end{document}